\documentclass[aps,prl,twocolumn,groupedaddress,latterpaper]{revtex4}
\usepackage{graphicx}

\begin{document}

\title{Field and intensity correlations in amplifying random media}
\author{Alexey Yamilov$^1$, Shih-Hui Chang$^2$, Aleksander Burin$^1$, Allen Taflove$^2$, Hui Cao$^1$}
\affiliation{$^1$ Department of Physics and Astronomy\\$^2$ Department of Electrical and Computer Engineering\\ Northwestern University, Evanston, IL 60208}
\email{a-yamilov@northwestern.edu}

\date{\today}

\begin{abstract}
We study local and nonlocal correlations of light transmitted through active random media. The conventional approach results in divergence of ensemble averaged correlation functions due to existence of lasing realizations. We introduce conditional average for correlation functions by omitting the divergent realizations. Our numerical simulation reveals that amplification does not affect local spatial correlation. The nonlocal intensity correlations are strongly magnified due to selective enhancement of the contributions from long propagation paths. We also show that by increasing gain, the average mode linewidth can be made smaller than the average mode spacing. This implies that light transport through a diffusive random system with gain could exhibit some similarities to that through a localized passive system, owing to dominant influence of the resonant modes with narrow width.
\end{abstract}
\pacs{42.25.Dd,42.25.Bs,42.25.Kb}


\maketitle

Without a counter-part in the electronic system, the coherent amplification of light adds a new dimension to the fundamental study of mesoscopic wave transport. An inherent quantum/wave signature of mesoscopic transport is nonlocal intensity correlation \cite{altshuler,nonlocal,feng_and_vanrossum}, which reflects the closeness to Anderson localization transition \cite{ping_sheng}.  Light transport in an amplifying random medium experiences  enhanced contribution from long paths \cite{zyuzin}, that should have a profound effect on the nonlocal intensity correlation.

Due to formal similarity, it is tempting to treat a random system with gain as if it had ``negative absorption'' \cite{duality}, and directly adopt the results obtained for an absorbing system. Such simplistic approach to correlation functions (CFs) is fundamentally flawed. Theoretically, the spatial and spectral CFs are obtained by average over an infinite number of random realizations. Among them, there exist rare configurations containing more localized modes that could lase in the presence of gain. Light intensity in the lasing configurations diverges, so do the ensemble-averaged CFs. Experimentally, the divergence of laser intensity is prevented by gain saturation. Nevertheless, the lasing configurations have much higher intensity than the non-lasing ones, thus they dominate the CFs. The width of spectral CFs is simply equal to the lasing mode linewidth, while the spatial CFs only reflect the spatial extent of the lasing modes. This is in contrast to the ``negative absorption'' model, which does not contain the divergent contribution of the rare events \cite{burkov}. In order to obtain the CFs that reflect light transport in amplifying random media, we introduce the conditional average over all non-lasing configurations $\langle ...\rangle \rightarrow \langle ...\rangle_c$. Such replacement, together with the fact that the fraction of lasing configurations varies with the amount of gain, make any analytical derivation challenging. Numerical simulations turned out to be a fruitful alternative. 

In this paper, based on numerical simulations, we present a phenomenological analysis of the local and nonlocal correlations of light transmitted through active random media. The systems under consideration are in the diffusive regime, but not too far from the localization threshold. We show that ``negative absorption'' formulae give a good fit to the conditional CFs only at low gain, with a decrease of the dimensionless conductance $g$ this range of applicability is further reduced. For high gain, even after discarding the contributions of lasing configurations, the long-range correlations are significantly stronger than the prediction of ``negative absorption'' theory, especially for the systems closer to the localization threshold. This is surprising in the sense that removal of all lasing configurations, which are dominated by more localized modes and have stronger nonlocal correlations, should have weakened the long-range correlations averaged over the rest non-lasing (less localized) configurations. Moreover, the average mode linewidth $\delta \nu$, found from the width of conditional spectral CF, does not exhibit any widening compared to the ``negative absorption'' expression. This is unexpected because exclusion of the narrowest modes (which have lased) should have overestimated the ``average'' mode linewidth. Therefore, the enhancement of long-range correlations and narrowing of spectral correlation width caused by coherent amplification exceed the expectations from ``negative absorption'' model. It reveals the absence of duality between gain and absorption. We also calculate the effective Thouless number $\delta \equiv  \delta \nu / d \nu$, where $d \nu$ is the average mode spacing. In the absence of gain, the onset of localization is marked by $\delta = 1$. We show that for diffusive systems, as gain increases, $\delta$ decreases monotonically to below 1 before reaching the diffusive lasing threshold predicted by the ``negative absorption'' theory that ignores fluctuations. This is an intriguing result, which seems to imply that transport in a diffusive system with gain could exhibit some similarities to that through a localized passive system. 

\begin{figure}
\centerline{\rotatebox{0}{\scalebox{0.4}{\includegraphics{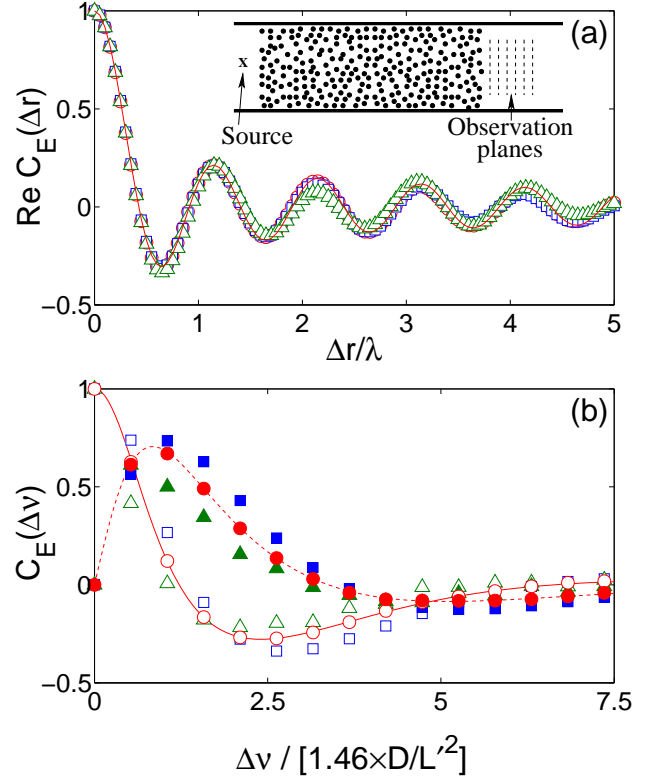}}}}
\caption{\label{ce_dr} (a) Real part of $C_E(\Delta r)$, (b) real (open symbols, solid line) and imaginary (solid symbols, dashed line) parts of $C_E(\Delta \nu)$. Circles, squares, and triangles represent numerical data for passive ($\tau_{abs/amp} = \infty$), absorbing ($\tau_{abs} = -5 \, \tau_{amp}^{cr}$), and amplifying ($\tau_{amp} = 5 \,\tau_{amp}^{cr}$) systems, respectively.  $\tau_{amp}^{cr}$ is the critical amplification time constant at the diffusive lasing threshold. $L$=0.4 m, $W$ = 0.2 m, $l$ = 0.18 cm. Curves represent theoretical fit with $z_b/l=0.8$. The inset is a sketch of the numerical experiment.}
\end{figure}

In our numerical simulation, we consider 2D random medium in a waveguide geometry, shown in the inset of Fig. \ref{ce_dr}a. It consists of a metallic waveguide filled with circular dielectric scatterers of refractive index $n=2$. Our numerical method for calculation of CFs in passive systems has been described elsewhere \cite{our_passive_work}. Physically, our system is quasi one-dimensional, and the transition from diffusion to localization can be realized by increasing the length $L$ of the random medium. To demonstrate the independence of our results on microscopic structure of the system, we varied both the filling fraction of scatterers and length of the random medium to obtain samples with $g$ from 2.2 to 9. The effect of absorption or gain (inside the scatterers) is treated by classical Lorenzian model\cite{taflove} with positive or negative conductivity. The advantage of our numerical model is the ability to introduce spatially uniform gain as well as to separate coherent amplification of an input signal from spontaneous emission of the active medium. In the numerical experiment, a short pulse was launched via a point source at the input end of the waveguide. Fourier transform of the electromagnetic field at the output end gave the CFs for field\cite{shapiro_ce} $C_E(\Delta r, \Delta \nu)$  and intensity\cite{feng_and_vanrossum} $C(\Delta r, \Delta \nu)$. In the presence of gain, long after the short excitation pulse, the electromagnetic field decays with time in the non-lasing realizations, while it keeps increasing in the lasing ones.  We excluded the lasing realizations from the ensemble average for CFs.  

Based on the pairing of incoming and outgoing channels, three \cite{shapiro_c0} contributions to intensity CF have been identified\cite{feng_and_vanrossum}: a local (short-range) $C_1\approx|C_E|^2$, and two nonlocal $C_2$ (long-range)  and  $C_3$ (infinite-range) ones. For diffusive transport $g\gg 1$ in a wave-guide geometry  $C_1\sim 1$, $C_2\sim 1/g$ and $C_3\sim 1/g^2$, making the values of $C_2$ and $C_3$ small\cite{feng_and_vanrossum}. The nonlocal terms are brought about by long propagation paths which are most sensitive to the effect of amplification. Despite of their small values in the passive systems, we observe a dramatic enhancement of the nonlocal correlations by coherent amplification.

Spatial field CF in 3D bulk was originally found by Shapiro in Ref. \cite{shapiro_ce}. Later Eliyahu {\it et al}\cite{eliyahu_surf} calculated it at the output surface of 3D random medium. Similarly, we derived the corresponding expression in 2D case\cite{our_passive_work}:
\begin{equation}
C_E(\Delta r)=\frac{\pi (z_b/l) J_0(k\Delta r) +2\sin (k\Delta r)/k\Delta r}{\pi z_b/l+2},
\label{ce_dr_theory}
\end{equation}
where $k=2\pi/\lambda$, $l$ is the transport mean free path, $J_0$ is Bessel function of zero order, and the extrapolation length $z_b\sim l$ accounts for boundary effects. The imaginary part of $C_E(\Delta r)$ should vanish due to isotropy\cite{genack_symmetry}, which is confirmed by our calculation where its value is less than $10^{-3}$. The real part of $C_E(\Delta r)$ are found unchanged in the presence of gain or absorption, as shown in Fig. \ref{ce_dr}a for a system of $g$ = 2.2. Eq. (\ref{ce_dr_theory}) gives excellent fit for passive, absorbing, and amplifying systems with the same value of $z_b/l$. Physically, this invariance can be explained by the local nature of $C_E(\Delta r)$. Spatial field correlation contains information that comes from the length scale of mean free path. $l$ is always shorter than the ballistic gain length $l_g$: $l_g/l>(2n_{eff}^{(e)}/\pi^2)\cdot(L/l)^2\gg1$, because the system is below the diffusive lasing threshold $L< \pi l_{amp}$. In the above expressions, the amplification length $l_{amp} = \sqrt{D \tau_{amp}}$, where $D$ is the diffusion coefficient and $\tau_{amp} = l_g/c$, the effective index of refraction $n_{eff}^{(e)}=c/v_E$, where $v_E$ is the energy transport velocity. Since amplification occurs on the scale much longer than $l$, it has negligible effect on short-range transport and local spatial correlations.  

\begin{figure}
\centerline{\rotatebox{-90}{\scalebox{0.32}{\includegraphics{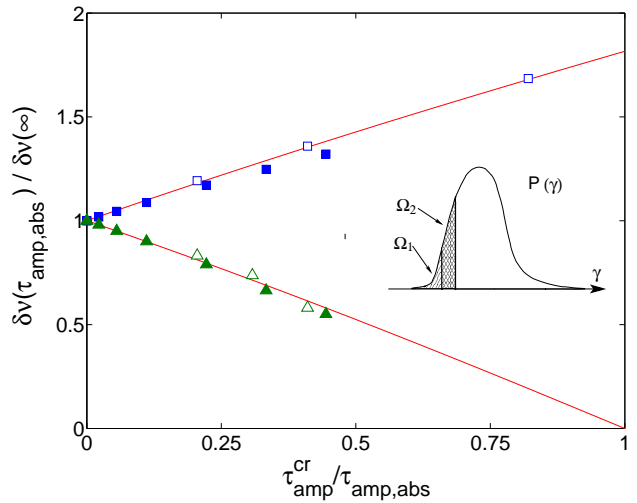}}}}
\caption{\label{dnu} Spectral correlation width $\delta \nu$ as a function of amplification time $\tau_{amp}$ (triangles) or absorption time $\tau_{abs}$ (squares). $\delta \nu$ is normalized to the value without gain or absorption ($\tau_{amp/abs}=\infty$), $\tau_{amp/abs}$ to $\tau^{cr}_{amp}$. Solid symbols correspond to $g = 4.4$, open symbols to $g=9$. The solid curves are given by the diffusion theory. The inset schematically shows the distribution of the resonant mode linewidth, see text for discussion.}
\end{figure}

The spectral field CF $C_E(\Delta \nu)$ contains an important dynamical parameter of transport - the diffusion coefficient $D=v_E\ l/2$. The spectral correlation width $\delta \nu$ is defined as the width at half maximum of $|C_E(\Delta \nu)|^2$ divided by a numerical factor $1.46$. In a passive system, $\delta \nu$ is equal to the average mode linewidth $D/L^{\prime\ 2}$, where $L^{\prime} = L+2z_b$. Since $v_E$ can be determined separately through calculation of energy distribution between air and dielectric scatterers\cite{our_passive_work}, the transport mean free path was found by fitting of the real and imaginary parts of $C_E(\Delta \nu)$ [Fig. \ref{ce_dr}b]. The value of $l$ allowed us to determine $g=(\pi/2) n_{eff}^{(e)}Nl/L^\prime$, where $N=2W/\lambda$ is the number of propagating modes in the waveguide, and $W$ is its width. In the presence of absorption\cite{garciaANDkogan}, the numerically calculated $C_E(\Delta \nu)$ fits well the expression derived in Refs. \cite{pniniANDvan_rossum}. For the case of amplification, we  obtained the ``negative absorption'' formula by making the substitution $-\tau_{abs} \rightarrow \tau_{amp}$:
\begin{equation}
C_E(\Delta \nu)=\frac{\sinh(q_0 a)}{\sinh(q_0 L^\prime)}\frac{\sin(L^\prime/l_{amp})}{\sin(a/l_{amp})},
\label{ce_dnu}
\end{equation}
where $q_0=\gamma_+-i\gamma_-$, $\gamma_\pm^2=\left(\sqrt{1/l_{amp}^4+\beta^4}\mp 1/l_{amp}^2\right)/2$, $\beta=\sqrt{2\pi\Delta\nu/D}$, and $a\simeq l$ is randomization length. By fitting $C_E(\Delta \nu)$ with Eq.(\ref{ce_dnu}), we obtained $\delta \nu$, which is plotted in Fig. \ref{dnu} for systems of $g=4.4$ and $9.0$. The narrowing of spectral correlation width by gain is due to partial compensation of light leakage through the system boundaries. Absorption, on the contrary, introduces an additional loss mechanism,  that leads to an increase of $\delta \nu$.  For both amplifying and absorbing media, the calculated $\delta \nu$ agrees well with the diffusion prediction. This agreement is surprising for the case of amplification. The ``negative absorption'' theory neglects the fluctuation of lasing threshold, and assumes the spectral width of all modes decreases with gain uniformly. However, the width $\gamma$ of resonant modes has a distribution $P(\gamma)$, schematically plotted in the inset of Fig. \ref{dnu}. For a given amount of gain, the modes with small $\gamma$ in the tail ($\Omega_1$) of $P(\gamma)$ lase, and they are excluded from the ensemble average. Such selective elimination of the narrowest modes should have led to an overestimation of $\delta \nu$. The absence of deviation from Eq.(\ref{ce_dnu}) indicates amplification not only reduces the width of all non-lasing modes, but also enhances the weight of the modes with narrower-than-average width (in $\Omega_2$) in the averaging.  

In Fig. \ref{dnu}, the diminishing correlation width in amplifying system signifies approach to lasing threshold for the mode with average linewidth. According to Eq. (\ref{ce_dnu}), $\delta \nu = 0$ when $\sin{(L^\prime/l_{amp})}$ turns to zero at $L^\prime/l_{amp}=\pi$. This ``average'' lasing threshold agrees with the diffusive lasing threshold derived by Letokhov\cite{letokhov}. Our calculation shows that the (conditional) average mode linewidth $\delta \nu$ can become smaller than the average mode spacing $d \nu$ before the diffusive lasing threshold is reached. Namely, with increasing gain, $\delta \nu$ decreases to $d \nu$ before reaching zero. This means the effective Thouless number $\delta$ can be reduced to below 1 by coherent amplification for a system that is diffusive in the absence of gain.

\begin{figure}
\centerline{\rotatebox{-90}{\scalebox{0.35}{\includegraphics{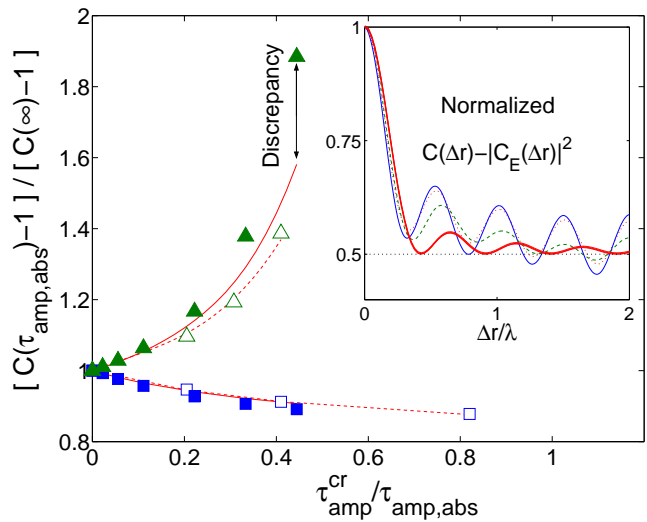}}}}
\caption{\label{c_long} $C-C_1$ at $\Delta r=0$ and $\Delta \nu=0$ in absorbing (squares) and amplifying (triangles) systems. Solid symbols correspond to $g=4.4$, open symbols to $g=9$. Solid and dashed curves are obtained from Eq.(\ref{c2pc3}) without any fitting parameters. The inset compares the dependence of $C-C_1$ on $\Delta r$ with $(F(\Delta r)+1)/2$ (thick line). Thin solid, dotted, and dashed lines represent passive, absorbing, and amplifying systems as in Fig.\ref{ce_dr}. }
\end{figure}

Fig. \ref{c_long} shows the nonlocal part of spatial intensity CF, $C(\Delta r,\tau^{cr}_{amp}/\tau_{amp})-|C_E(\Delta r,\tau^{cr}_{amp}/\tau_{amp})|^2$, in samples of $g=4.4$ and $9.0$. According to Refs. \cite{bing_hu,seba_book}, spatial variation and absorption contribution should factorize. Accounting for terms up to $1/g^2$ we obtained the ``negative absorption'' expression for nonlocal intensity correlation \cite{garciaANDkogan,pniniANDvan_rossum,seba_book,andrey_polarization,brouwer}
\begin{eqnarray}
C(\Delta r,s=L/l_{amp})-|C_E(\Delta r,s)|^2=\left( 1+F(\Delta r)\right)\times\ \ \ \ \  
\label{c2pc3}
\\
\left[\frac{1}{4gs}\frac{2s(2-\cos{2s})-\sin{2s}}{\sin^2{s}}+\frac{4}{g^2}\frac{\sin^2{s}}{s^2}\times \right.  \ \ \ \ \ \nonumber
\\
\left. \left(\frac{2s^2-s\cot{s}+1}{16\sin^2{s}}-3\frac{s^2+s\cot{s}+1}{16\sin^4{s}}+\frac{3s^2}{8\sin^6{s}}\right)\right],\nonumber
\end{eqnarray}
where $F(\Delta r)=|C_E(\Delta r)|^2$. The inset of Fig. \ref{c_long} plots the profile of $C(\Delta r)-C_1(\Delta r)$, normalized to its value at $\Delta r = 0$. For passive, absorbing and amplifying systems, the dependence of $C-C_1$ on $\Delta r$ are almost the same. In particular, the value of $C-C_1$ at $\Delta r\rightarrow \infty$ is exactly 1/2 of that at $\Delta r=0$, in agreement with $[ 1+F(\Delta r)]$ dependence. This tells us amplification (absorption) increases (decreases) the nonlocal correlations at every $\Delta r$ uniformly. Therefore, the enhancement (reduction) can be characterized by a single number, e.g., the value of $C-C_1$ at $\Delta r=0$ as shown in Fig. \ref{c_long}. In two absorbing samples of $g=4.4$ and $9$, the decrease of nonlocal correlations is in good agreement with the diffusion prediction. For amplifying media, only when the fraction of omitted lasing realizations is small, Eq.(\ref{c2pc3}) adequately describes the nonlocal correlations of the transmitted intensity. For high gain, we see strong deviations: even after removing the lasing realizations, nonlocal correlation still exceeds the ``negative absorption'' prediction [Eq. (\ref{c2pc3})]. The deviation becomes more pronounced as the dimensionless conductance decreases. The rapid  increase of nonlocal correlation with gain is caused by enhanced contribution from long trajectories that cross upon themselves. 

\begin{figure}
\centerline{\rotatebox{-90}{\scalebox{0.33}{\includegraphics{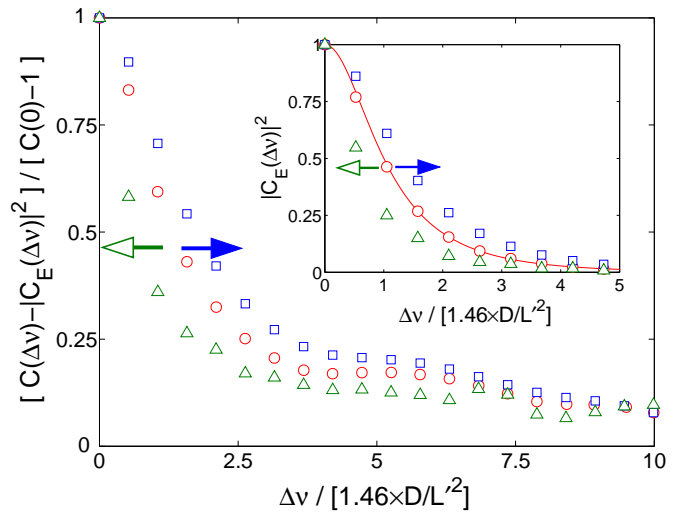}}}}
\caption{\label{c_long_vs_dnu} Frequency dependence of nonlocal contribution to $C(\Delta \nu)$ normalized to its value at $\Delta \nu=0$. The inset shows the local contribution $C_1(\Delta \nu)=|C_E(\Delta \nu)|^2$. System parameters and symbol notations are the same as in Fig. \ref{ce_dr}. The open (solid) arrow shows the direction of increasing gain (absorption).}
\end{figure}

Finally, we calculated the spectral correlations of transmitted intensities. Fig. \ref{c_long_vs_dnu} reveals the changes of $C_1(\Delta \nu)$ and $C(\Delta \nu)-C_1(\Delta \nu)$ due to gain and absorption. The frequency separation $\Delta \nu$ is normalized to the average mode linewidth $\delta \nu$ in the absence of gain or absorption. $C(\Delta \nu)-C_1(\Delta \nu)$ is also normalized to its value at $\Delta \nu = 0$. Both $C_1$ and $C-C_1$ are narrowed by gain, and broadened by absorption. While the narrowing of $C_1(\Delta \nu)$ with increasing gain is similar to that of a passive system with decreasing $g$, the narrowing of $C(\Delta \nu)-C_1(\Delta \nu)$ is just the opposite\cite{our_passive_work}. When a passive system approaches the localization transition, the most conducting channels dominate the transport and correlations. Thus, after normalizing $\Delta \nu$ by the correlation width $\delta \nu$, $C(\Delta \nu)-C_1(\Delta \nu)$ is widened as $g$ decreases in a passive system, but it is narrowed with the fixed $g$ and increasing gain.  This remarkable difference illustrates that in a passive system near the localization transition, the modes with larger-than-average width dominate the transport; while in a diffusive system with gain, the modes with narrower-than-average width become dominant. 

This work is supported by the National Science Foundation under Grant No DMR-0093949. HC acknowledges the support from the David and Lucille Packard Foundation.

\end{document}